\documentclass[a4paper,twocolumn,11pt, accepted=2023-12-14]{quantumarticle} 
\pdfoutput=1
\usepackage[utf8]{inputenc}
\usepackage[english]{babel}
\usepackage[T1]{fontenc}
\usepackage{float}
\usepackage{amsmath, amsfonts,physics}
\usepackage{xcolor}
\usepackage{hyperref}
\hypersetup{
    colorlinks=true,
    citecolor=blue,
    urlcolor=violet,
    linkcolor=teal,
    }

\usepackage[shortlabels]{enumitem}

\usepackage[numbers]{natbib} %Unicode characters in BibTeX

\usepackage{booktabs} % Horizontal rules in tables

\usepackage{physics}

\usepackage{tikz}
\usepackage{lipsum}

\usepackage{graphicx}
\usepackage{tikz}
\usetikzlibrary{arrows,positioning,calc, shapes,backgrounds,fit}  
\usetikzlibrary{arrows.meta}
\usetikzlibrary{decorations.text}
\usetikzlibrary{decorations}
\usetikzlibrary{%
    decorations.pathreplacing,%
    decorations.pathmorphing%
}

\usepackage{dcolumn}% Align table columns on decimal point
\usepackage{bm}% bold math

\usepackage{dsfont} %to include identity operator \mathds{1}
\usepackage{booktabs}
\usepackage{float}

\usepackage{tikz}
\usetikzlibrary{decorations.text}
\usetikzlibrary{decorations}
\usetikzlibrary{%
    decorations.pathreplacing,%
    decorations.pathmorphing%
}
\usepackage[ruled, lined, linesnumbered, commentsnumbered, longend]{algorithm2e}
\newcommand{\correction}[1]{\textcolor{black}{#1}}

\begin{document}

\title{Determining the ability for universal quantum computing: Testing controllability via dimensional expressivity}

\author{Fernando Gago-Encinas}
\affiliation{%
Fachbereich Physik and Dahlem Center for Complex Quantum Systems, Freie Universit\"{a}t Berlin, Arnimallee 14, 14195 Berlin, Germany
}%
\author{Tobias Hartung}%
\affiliation{%
Northeastern University London, Devon House, St Katharine Docks, London, E1W 1LP, United Kingdom
}%
\affiliation{%
Khoury College of Computer Sciences, Northeastern University, 440 Huntington Avenue, 202 West Village H
Boston, MA 02115, USA
}%
\author{Daniel M. Reich}
\affiliation{%
Fachbereich Physik and Dahlem Center for Complex Quantum Systems, Freie Universit\"{a}t Berlin, Arnimallee 14, 14195 Berlin, Germany
}%
\author{Karl Jansen}%
\affiliation{%
NIC, DESY Zeuthen, Platanenallee 6, 15738 Zeuthen, Germany
}%

\author{Christiane P. Koch}
\affiliation{%
Fachbereich Physik and Dahlem Center for Complex Quantum Systems, Freie Universit\"{a}t Berlin, Arnimallee 14, 14195 Berlin, Germany
}%

\date{December 15, 2023}

\begin{abstract}
Operator controllability refers to the ability to implement an arbitrary unitary in $SU(N)$ and is a prerequisite for universal quantum computing. Controllability tests can be used in the design of quantum devices to reduce the number of external controls. Their practical use is hampered, however, by the exponential scaling of their numerical effort with the number of qubits. Here, we devise a hybrid quantum-classical algorithm based on a parametrized quantum circuit. We show that controllability is linked to the number of independent parameters, which can be obtained by dimensional expressivity analysis. We exemplify the application of the algorithm to qubit arrays with nearest-neighbour couplings and local controls.  Our work provides a systematic approach to the resource-efficient design of quantum chips. 
\end{abstract}

\maketitle

%------------------------------------------------
 
\section{Introduction}

Universal quantum computing~\cite{nielsen2010quantum} requires controllability on the quantum processing unit, so that every quantum logic gate can be implemented. A common layout in hardware platforms such as those based on superconducting circuits achieves this by combining two-qubit couplings with local drives for each qubit of the array~\cite{krantz2019quantum, garcia2022quantum}. 
While effective, this approach becomes demanding for larger arrays, due to both the physical space needed for each control as well as the associated calibration. Controllability tests can help identify less resource-intensive architectures that are still capable of performing the same quantum gates~\cite{gago2023graph}. 

Controllability in general studies the dynamics that can be implemented in a quantum system driven by a set of controls~\cite{dAlessandro2021,KochEPJQT22, GlaserEPJD15}. In particular, a system is pure-state controllable if it can reach all final states. Alternatively, an (evolution) operator controllable system is capable of implementing every unitary gate, a necessary feature for universal quantum computing. Tests for these two different types of controllability rely on computing the rank of the dynamical Lie algebra of the Hamiltonian~\cite{dAlessandro2021} or utilize methods based graph theory~\cite{AlbertiniLinAlg2002, BCCS, boscain2014multi, gago2023graph}. For small system sizes, the tests can be carried out analytically~\cite{SchirmerPRA2001,FuJPhysA2001,AltafiniJMP2002, pozzoli2022}.  For some high- and infinite-dimensional systems, controllability can be determined using induction arguments~\cite{chambrion2009,Boussaid2013,BCCS,leibscher2022}. Beyond these special cases, a numerical approach is possible in principle~\cite{gago2023graph}, but is limited by the exponential scaling of the Hilbert space dimension with respect to the number of qubits. In other words, the accuracy and feasibility of controllability tests for increasing system size suffer from the curse of dimensionality. 

Here, we present a hybrid quantum-classical controllability test, for both pure-state and operator controllability of qubit arrays. The hybrid method we propose evaluates the controllability of the qubit array by measurements on a quantum device, either the system to be studied with an extra \correction{auxiliary} qubit or one that mimics the dynamics of the original system. This opens up a new way of designing controllable qubit arrays with fewer resources, helping to address the issue of scalability. To do so, we harness the computational power of quantum circuits to extract information directly from the qubit array under study. \correction{While our controllability test relies on the same mathematical foundations as the Lie rank condition, its design circumvents issues that arise in the construction of the dynamical Lie algebra, in particular the orthonormalization calculations it entails. Mapping the classical operations of our hybrid quantum-classical algorithm to a quantum device would further expand the size of the qubit arrays whose controllability we can determine.}

Parametric quantum circuits constitute the basis of many algorithms, for example variational algorithms for solving computationally hard optimization problems~\cite{peruzzo2014variational, mcclean2016theory}.
The circuits consist of a set of parametric gates that can be used to measure a cost function. After a classical optimization, the parameters are updated to give a new cost value, continuing the feedback loop of the algorithm. It is necessary to include enough independent optimization parameters to reach the best possible solution. 
However, minimizing the number of parametric gates and circuit depth is also key in the era of noisy quantum devices~\cite{preskill2018quantum}. In order to reduce the noise of the circuit while maintaining its optimization capability, every redundant parameter should be identified and removed from the circuit. This goal is related to the dimensional expressivity of the circuit and can be achieved with dimensional expressivity analysis~\cite{funcke2021DEA, funcke2021DEA_approx}, a hybrid quantum-classical algorithm to systematically find redundant parameters. 

In order to leverage dimensional expressivity analysis to test for controllability, we define a parametric quantum circuit based on the architecture of a given qubit array with local controls and qubit couplings. We then use dimensional expressivity analysis to quantify the number of independent parameters which is related to the controllability of the original qubit array. We provide a complete description of how to carry out the hybrid controllability test on a quantum circuit, opening the possibility of obtaining information of the controllability of a quantum device before it is built.

The manuscript is organized as follows. The basic concepts of controllability analysis and parametric quantum circuits are briefly reviewed in section~\ref{sec:theo_background}. The pure-state controllability test is presented in section~\ref{sec:PSC_controllability}, including its derivation, definition and showcase examples. Section ~\ref{sec:OC_controllability} extends the test to operator controllability, making use of the Choi-Jamio\l kowski isomorphism. Section~\ref{sec:conclusions} concludes.

%------------------------------------------------

\section{Theoretical background}\label{sec:theo_background}

To define controllability tests for qubit arrays, we combine the notions of system controllability and circuit expressivity. For the sake of a self-contained presentation, we briefly recap the basic concepts in this section. 

\subsection{Controllability}

We consider quantum systems linearly coupled to external controls. They are described by traceless Hamiltonians of the form
\begin{equation}
    \hat{H}(t) = \hat{H}(t; u_1, ... u_m) = \hat{H}_0 + \sum_{j=1}^m u_j (t) \hat{H}_j, \label{eqn:controlH}
\end{equation}
where $u_j (t)$ are the controls and $\hat{H}_j$ are the control operators.
%The evolution of any state $\ket{\psi (0)}$ in the Hilbert space $\mathcal{H}$ of the system under Eq.~\eqref{eqn:controlH} is given by $\ket{\psi (t)} = \hat{U}(t; u_1, ... u_m) \ket{\psi (0)}$.
The Hamiltonian \eqref{eqn:controlH} generates the time evolution operator $\hat{U}(t)$ such that for any state $\ket{\psi(0)}$ in the Hilbert space $\mathcal{H}$, $\ket{\psi(t)} = \hat U(t) \ket{\psi (0)}.$
Given an initial state $\ket{\psi_0}$, the set of all final states $\ket{\psi (T)}$ that can be reached in finite time $0<T<\infty$ with controls $u_j (t)$ is called the reachable set $\mathcal{R}_{\ket{\psi_0}}$ of the system. The system is said to be pure-state controllable if $\mathcal{R}_{\ket{\psi_0}}=\mathcal{S}^{\mathcal{H}}$ (with $\mathcal{S}^{\mathcal{H}}$ the unit sphere on $\mathcal{H}$), i.e., if every normalized state is reachable from any initial state $\ket{\psi_0}$~\cite{dAlessandro2021}. For physical systems this condition is not dependent on the initial state, which means that pure-state controllability is also independent of the state in which the system is initialized. Indeed, the evolution operators $\hat{U}$ that can be implemented on such systems form a group. This implies that for every evolution $\hat{U}$ in the group, $\hat{U}^{-1}$ must also be contained in the group of feasible evolutions. If we assume that every state $\ket{\phi}\in \mathcal{H}$ can be reached from a certain initial state $\ket{\psi_0}$, then for every state $\ket{\phi}$ there exists an evolution $\hat{U}_{\ket{\psi_0}, \ket{\phi}}$ such that $\ket{\phi} = \hat{U}_{\ket{\psi_0}, \ket{\phi}} \ket{\psi_0}$. Therefore given any initial state $\ket{\phi_i}$ and final state $\ket{\phi_f}$ we can always generate an evolution
\begin{equation}
   \ket{\phi_f} =  \hat{U}_{\ket{\psi_0}, \ket{\phi_f}} \hat{U}^{-1}_{\ket{\psi_0}, \ket{\phi_i}}\ket{\phi_i}
\end{equation}
In particular this proves that if all states are reachable from a certain initial state in a closed system, every state is reachable from any other state. 

Pure-state controllability is the relevant type of controllability when we are interested in are state transfers, i.e., evolving the system from an initial state to a certain target state. It is equivalent to proving that all state transfers are possible in a system. This, however, is not the strongest type of controllability that can be defined. Pure-state controllability is sufficient to guarantee that there will always be evolution operators $\hat{U}_{\ket{\psi_0}, \ket{\psi_f}}$ to connect any two states $\ket{\psi_0}$ and $\ket{\psi_f}$, yet not enough to ensure that it is possible to generate every operation $\hat{U}$ in the special unitary group $SU\left(d\right)$, where $d = \dim (\mathcal{H})$. Pure-state controllability does not guarantee that simultaneous state-to-state transfers are always possible. To study this property we consider the so-called operator controllability. A system with controls as defined in \eqref{eqn:controlH} and Hilbert space dimension $d$ is operator controllable if for every unitary evolution $\hat{U}_{target} \in SU(d)$ there exist a final time $T\geq 0$, a phase angle $\varphi \in [0,2\pi)$ and a set of controls $\{u_j\}_{j=1}^m$ such that $\hat{U}_{target} =  e^{i \varphi} \hat{U}(T; u_1, ... u_m)$. 

Note that for both types of controllability there are no restrictions on the final time $T\leq \infty$ at which state transfers, respectively unitary operations, are implemented. Consequently, this time $T$, while always finite, can be arbitrarily large. The question of controllability only inquires whether it is possible at all to perform the desired dynamics. Similarly, it does not impose any restrictions on the maximum amplitude that the controls $u_j (t)$ from \eqref{eqn:controlH} can take. Finite amplitude is a physical restriction that impacts the final time required to perform the different operations, but does not mathematically change the controllability of the system. 

If the Hamiltonian of the system is known, there exist algebraic and numerical tests tailored for both types of controllability~\cite{altafini2002controllability, albertini2003notions, schirmer2002identification, dAlessandro2021, gago2023graph}. 

%    ----    ----    ----    ----    ----    ----    ----

\subsection{Dimensional expressivity}

Parametric quantum circuits have multiple applications, as they constitute the base for variational quantum algorithms~\cite{cerezo2021variational}. Their design and study are pivotal factors in the efficiency of the algorithms. In particular, parameter dependence and the set of final states that can be produced are two key topics that determine the capability of the algorithms. Lacking some necessary parametric gates leads to unsuccessful algorithms, whereas including too many dependent parameters is detrimental for the purpose of optimization. We introduce here notions and definitions related to these issues that are relevant for the controllability tests. 

\begin{center}
\begin{figure}[tb]
\includegraphics[width=0.45\textwidth]{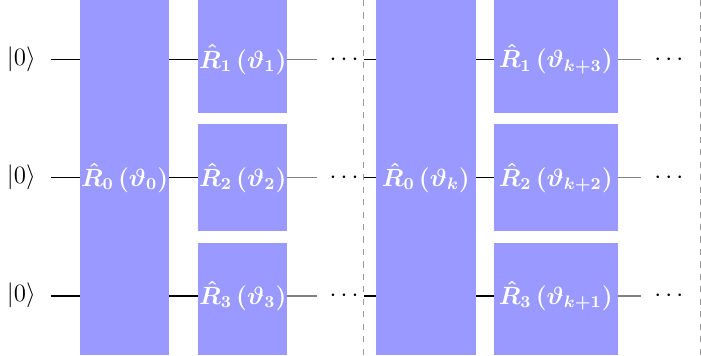}
\caption{\label{fig:PSC_circuit}  Three-qubit example of the parametric circuit $C_{PSC}(\vec{\vartheta})$ (\ref{eqn:PSC_test}) for testing pure-state controllability with initial state $\ket{000}$ in the qubits' logical basis. Each layer (only two displayed in the diagram) includes an entangling gate $\hat{R}_0$ and a sequence of local gates $\hat{R}_j$ (with $j\geq 1$), one for every control present in the qubit array. }
\end{figure}
\end{center}

A parametric quantum circuit is a protocol implemented on a set of qubits that are initialized in a state $\ket{\psi_0}$. It consists of a sequence of logic gates $\hat{G}_j$, some of which depend on real parameters $\vartheta_k$. We consider a parametric quantum circuit as the map $C(\vec{\vartheta})$ that identifies an array of parameters $\vec{\vartheta}$ in the parameter space $\mathcal{P}\ni \Vec{\vartheta}$ with  
\begin{equation}\label{eqn:PQC}
    C(\vec{\vartheta}) = \hat{G}_m (\vec{\vartheta}) ...\hat{G}_0 (\vec{\vartheta}) \ket{\psi_0}. 
\end{equation}
$C(\vec{\vartheta})$ implicitly depends on the circuit's initial state $\ket{\psi_0}$\footnote{This is in contrast to many commonly used definitions of a circuit that only consider the gate sequence and not the device initialization. }. An example of a parametric quantum circuit is found in Figure~\ref{fig:PSC_circuit}. Note that the amount of parameters on which each gate $\hat{G}_j (\vec{\vartheta})$ depends may vary from zero to the total number of parameters, e.g.
\begin{equation}
    \hat{G}_0 (\vartheta_1, \vartheta_2) = \hat{P}(\vartheta_1)\exp\left(-i \frac{\vartheta_2}{2}\hat{X}\right) \hat{H}\hat{P}(-\vartheta_1),
\end{equation}
with the phase gate $\hat{P}$ and the Hadamard gate $\hat{H}$. For the sake of simplicity, we have chosen units such that $\hbar = 1$.  

The expressivity of a parametric quantum circuit is its ability to produce states that are representative of the full Hilbert space of the system~\cite{sim2019expressibility, friedrich2023quantum}. Here, we focus on the dimensional expressivity $expr_{dim}$, i.e. the dimension of $C\left(\mathcal{P}\right)$ as a real differentiable manifold~\cite{funcke2021DEA}. 
As such, the maximal dimensional expressivity for a circuit with complex Hilbert space dimension $d$ is $\max(expr_{dim}) = 2d-1$, which accounts for the real variables of the complex $d$-dimensional Hilbert space minus the normalization constraint. 

Another important point is the concept of redundant parameters. In a quantum circuit $C(\vec{\vartheta})$, a parameter $\vartheta_j$ is considered redundant if small perturbations on $\vartheta_j$ produce final states on $C(\vec{\vartheta})$ that can also be achieved by keeping $\vartheta_j$ constant and varying the rest of the parameters $\vartheta_k$ as needed~\cite{funcke2021DEA_approx}. Minimizing the number of redundant parameters is therefore a relevant matter in the design of parametric quantum circuits. Fewer redundant parameters may result in more resource-efficient circuits that can produce the same manifold of states. If a parameter $\vartheta_1$ is redundant with another parameter $\vartheta_2$, then the converse is also true. We are free to choose one of the two parameters to remain constant while varying the other one at will. The latter is then called independent. Mathematically, the dimensional expressivity of a circuit $C(\vec{\vartheta})$ is also equal to the number of elements in the maximal set of independent parameters in the circuit. While the cardinality of the maximal set for a certain circuit $C(\vec{\vartheta})$ is fixed, there may exist multiple maximal sets. Locating and eliminating redundant parameters results in a minimal circuit with the same local dimension in the manifold of reachable states. 

Redundant parameters and dimensional expressivity are studied through the real Jacobian $J_{C}$ of $C(\vec{\vartheta})$. Assuming a total of $N$ parameters, it takes the form
\begin{equation}
\arraycolsep=2.5pt
J_{C} (\vec{\vartheta})=   \left( \begin{array}{ccc} \label{eqn:Jacobian}
        | & & | \\
         \mathfrak{Re} \partial_1 C (\vec{\vartheta}) & \cdots &  \mathfrak{Re} \partial_N C (\vec{\vartheta}) \\
        | & & | \\
         & \,  &  \\
        | & & | \\
         \mathfrak{Im} \partial_1 C (\vec{\vartheta}) & \cdots  &  \mathfrak{Im} \partial_N C (\vec{\vartheta}) \\
        | & & | 
    \end{array} \right),
\end{equation}
where the elements $\partial_k C$ represent the partial derivatives of $C$ with respect to $\vartheta_k$. \correction{$\mathfrak{Re} \partial_k C$ and $\mathfrak{Im} \partial_k C $ denote the real and the imaginary part of $\partial_k C$, respectively.}
By definition, the dimensional expressivity is equal to the rank of $J_{C} (\vec{\vartheta})$. In terms of $J_{C}$, a parameter $\vartheta_j$ is redundant with respect to the other parameters $\{\vartheta_i\}_{i\neq j}$ at a point $\vec{\vartheta}$ if the $j$-th column of $J_{C} (\vec{\vartheta})$ is linearly dependent with respect to the set of all the other columns of $J_{C} (\vec{\vartheta})$, i.e. if the rank of $J_{C} (\vec{\vartheta})$ as a matrix remains the same after removing the $j$-th column. 

A systematic approach, for an ordered array of parameters $\Vec{\vartheta}$, relies on the partial real Jacobians $J_{C, n} (\vec{\vartheta})$
\begin{equation}
\arraycolsep=2.5pt
J_{C, n} (\vec{\vartheta})=   \left( \begin{array}{ccc} \label{eqn:partialJacobian}
        | & & | \\
         \mathfrak{Re} \partial_1 C (\vec{\vartheta}) & \cdots &  \mathfrak{Re} \partial_n C (\vec{\vartheta}) \\
        | & & | \\
         & \,  &  \\
        | & & | \\
         \mathfrak{Im} \partial_1 C (\vec{\vartheta}) & \cdots  &  \mathfrak{Im} \partial_n C (\vec{\vartheta}) \\
        | & & | 
    \end{array} \right),
\end{equation}
containing only the first $n$ columns of $J_{C} (\vec{\vartheta})$. If $\partial_1 C (\vec{\vartheta})\neq 0$ then $\vartheta_1$ is independent and we initialize the set of independent parameters as $\mathcal{N}_1 := \{\vartheta_1\}$; otherwise,  $\mathcal{N}_1 := \emptyset$. Then we can iterate over the following step. If $\rank (J_{C, k+1} (\vec{\vartheta})) > \rank (J_{C, k} (\vec{\vartheta}))$, then $\vartheta_{k+1}$ is independent and we update the set of independent parameters $\mathcal{N}_{k+1} = \mathcal{N}_{k} \cup \{\vartheta_k\}$. Else, $\vartheta_{k+1}$ is redundant and $\mathcal{N}_{k+1} = \mathcal{N}_{k}$. After all $N$ parameters have been checked, the set $\mathcal{N}_{N}$ is a maximal set of independent parameters and its cardinality is the dimensional expressivity of the circuit. The redundant parameters can be then removed from the circuit by setting them to a suitably chosen constant value.

The dimensional expressivity analysis follows this approach and provides an efficient method to find a maximal set of independent parameters on a quantum circuit~\cite{funcke2021DEA, funcke2021DEA_approx}. As a hybrid quantum-classical algorithm, it mixes measurements on the actual circuit and classical computations for the ranks. Instead of calculating the ranks of $J_{C, n}$, this method retrieves the entries of the matrices 
\begin{equation}\label{eqn:Smatrix}
    S_{C, n} (\vec{\vartheta})= J_{C, n}^T (\vec{\vartheta}) J_{C, n} (\vec{\vartheta}),
\end{equation}
which are $n\times n$ matrices whose rank equals the one of $J_{C, n} (\vec{\vartheta})$. The elements of $S_{C, n} (\vec{\vartheta})$ can be determined via measurements on the circuit with the inclusion of a single \correction{auxiliary} qubit, no matter the number of qubits in the original circuit~\cite{funcke2021DEA}.

%------------------------------------------------

\section{Pure-state controllability test using dimensional expressivity}\label{sec:PSC_controllability}

This section introduces the novel connection between the dimensional expressivity of quantum circuits and the pure-state controllability of quantum systems. We present the design of a circuit associated to a controlled system that allows us to check its pure-state controllability. We include two examples to showcase its functionality. 

\subsection{Circuit expressivity and pure-state controllability}\label{ssec:expr_and_control}

We consider a qubit array with Hamiltonian (\ref{eqn:controlH}). We identify the drift $\hat{H}_0$ as the time-independent part, which includes the local free-qubit Hamiltonians and some time-independent couplings between them. Similarly, the operators $\hat{H}_j$ with $1\leq j\leq m$ are coupled to the $m$ different external controls acting on the system. In order to use dimensional expressivity analysis to determine controllability of a qubit array, it is necessary to define a parametric quantum circuit that can be run on the system, according to the different controls at disposal. If we can show that all normalized states in the Hilbert space are reachable from a certain initial state using only gates generated by the system's controls, we have proven pure-state controllability. 

A straightforward choice for the possible parametric gates in the circuit is
\begin{equation} \label{eqn:Rgate_PSC}
   \hat{R}_{j} (\alpha):= \exp\left(-i \,\frac{\alpha}{2}   \hat{H}_j\right), \qquad 0\leq j\leq m,
\end{equation}
i.e. rotations around either the drift $\hat{H}_0$ or the control operators $\hat{H}_j$ (\ref{eqn:controlH}). The gates $\hat{R}_0 (\alpha)$ can be implemented by letting the system evolve under its time-independent drift  Hamiltonian $\hat{H}_0$ for a certain time $t = \frac{\alpha}{2}$. For the other gates, $\hat{R}_j (\alpha)$ with $j\geq 1$, we make use of the local controls. In these gates the $\hat{H}_0$ contribution can be neglected by assuming that the controls can be chosen such that $ \lVert u_j (t) \hat{H}_j \rVert \gg \lVert \hat{H}_0 \rVert $. A realistic approach to the $\hat{R}_j (\alpha)$ implementation is to consider short rotations with intense controls $u_j(t)$, so that the $\hat{H}_0$ contribution is insignificant in comparison. The amplitude of $u_j(t)$ is usually adjusted externally and it has no imposed restriction. 

We want to design a parametric quantum circuit $C_{PSC}(\Vec{\vartheta})$, starting with an arbitrary initial state $\ket{\psi_0}\in\mathcal{H}$ and exclusively composed of the rotation gates $\hat{R}_j (\vartheta_k)$. We then use dimensional expressivity analysis to measure the dimensional expressivity of the system. If it is maximal, i.e. $expr_{dim} = 2 d-1$ for $\dim(\mathcal{H}) = d$, we have a manifold of reachable states with local real dimension $2d-1$. This manifold is a subset of $\mathcal{H}$.
We now prove that it is in fact the whole unit sphere of $\mathcal{H}$. 
If we assume that the gates $\hat{R}_j (\alpha)$ are cyclic and that every parameter $\vartheta_k$ is used in a single rotation gate in the circuit, we can treat each $\vartheta_k$ as if it had periodic boundaries, i.e. $\vartheta_k \in \mathbb{S}^1$. For an array of $n$ parameters $\Vec{\vartheta}$ the parameter space verifies
\begin{equation}
    \mathcal{P} \cong \underbrace {\mathbb{S}^1  \times \cdots \times \mathbb{S}^1}_{n} \cong \mathbb{T}^n.
\end{equation}
This implies that $\mathcal{P}$ is a connected, compact set without boundary. Assume a circuit $C_{PSC}(\Vec{\vartheta})$ that has maximal dimensional expressivity. Then, the manifold of reachable states $C_{PSC}(\mathcal{P})\subseteq \mathcal{H}$ is a connected, compact manifold without boundary and with maximal local real dimension. Consequently $C_{PSC}(\mathcal{P}) = \mathcal{S}^{\mathcal{H}}\subset \mathcal{H}$. Thus, the system is pure-state controllable. 

So far, we have found a sufficient condition for pure-state controllability. We now want to identify a condition for non-controllable systems. To this end, we need to prove that there are some states that are not reachable by any of the possible dynamics that we can implement with the different operators $\hat{H}_j$ and their nested commutators. 
Hypothetically, we could do a sequence of the rotation gates \eqref{eqn:Rgate_PSC} around the drift, the control operators and their nested commutators and test if all of them are linearly independent. However, generating the exponential of the commutator of two control operators (or one control operator and the drift) $\exp{i\, \beta [\hat{H}_j, \hat{H}_k]}$ is no trivial task. It may require optimal control to generate a specific rotation for the exact angle $\beta$ and the chosen commutator $[\hat{H}_j, \hat{H}_k]$. Instead, we access the different commutators by concatenating a series of multiplications, as in the Baker-Campbell-Hausdorff formula: 
\begin{align}\label{eqn:BCH}
    \exp\left(i\, \alpha \hat{A}\right) &\exp\left(i\, \beta \hat{B}\right) =  \exp \left(i\alpha\hat{A} + i\beta\hat{B} \right.  \nonumber \\ 
    &- \frac{\alpha\beta}{2}[\hat{A}, \hat{B}] - \frac{i \,\alpha^2\beta }{12}[\hat{A},[\hat{A}, \hat{B}]]   \nonumber \\
    &+ \left.\frac{i\, \alpha\beta^2 }{12}[\hat{B}, [\hat{A}, \hat{B}]] \cdots \right).  
\end{align}
%where $\hat{A}$ and $\hat{B}$ are two Hermitian operators, e.g. the control operators $\hat{H}_j$ ($1\leq j \leq m$) or the drift $\hat{H}_0$ in the system to study. 
Assume that we have a parametric quantum circuit consisting of a sequence of $n$ rotations,
\begin{equation}\label{eqn:C_exp_seq}
    C_{seq}^{n}(\Vec{\vartheta}) := \exp\left(-i \,\vartheta_n   \hat{A}_n\right) \cdots \exp\left(-i \,\vartheta_1   \hat{A}_1\right) \ket{\psi_0}
\end{equation}
with $\hat{A}_j \in \{\hat{H}_k\}_{k=0}^m \, \forall 1\leq j \leq n$.
We can use Eq.~\eqref{eqn:BCH} multiple times on the exponential sequence on the right-hand side of Eq.~\eqref{eqn:C_exp_seq} to express it as a single exponential dependent on $\vec{\vartheta}$, the different operators ${A}_j$ and their nested commutators. Assume as well that the dimensional expressivity in the circuit $expr_{\dim} (C_{seq}^{n}(\Vec{\vartheta})) = d_{n}$ is less than the maximum possible. We define a new parametric circuit by adding one more rotation to the chain of operations,
\begin{equation}\label{eqn:concatenate_R}
    C_{seq}^{n+1}(\Vec{\vartheta}, \vartheta_{n+1}) := \exp\left(-i \,\vartheta_{n+1} \hat{A}_{n+1}\right) C_{seq}^{n}(\Vec{\vartheta}). 
\end{equation}
If the dimensional expressivity of $C_{seq}^{n+1}$ and $C_{seq}^{n}$ are the same for every $\vartheta_{n+1} \in \mathbb{R}$ and every $\hat{A}_{n+1} \in \{\hat{H}_k\}_{k=0}^m$, then the number of linearly independent $\partial_j C (\vec{\vartheta})$ remains the same. 
In other words, we are not able to find more linearly independent operators and thus, the dimensional expressivity of the system cannot be increased. This means that the manifold of reachable states does not have a maximal local dimension and hence there will be some states to which our initial state cannot evolve. Therefore the system is not pure-state controllable.  

There may be cases where, for given $C_{seq}^{n}(\Vec{\vartheta})$ and $\hat{A}_{n+1}$, there exist two different parameters $\vartheta_{n+1}$ and $\tilde{\vartheta}_{n+1}$ such that 
\begin{equation}
\begin{aligned}
    expr_{\dim} \left(C_{seq}^{n+1}\right. &\left.(\Vec{\vartheta},  \vartheta_{n+1})\right)  \\
    >&\;\;  expr_{\dim} \left(C_{seq}^{n+1}(\Vec{\vartheta}, \tilde{\vartheta}_{n+1}) \right).
\end{aligned}
\end{equation}
This is common in cases where $\tilde{\vartheta}_{j} = 0$ for every $1\leq j \leq n+1$. Looking at Eq.~\eqref{eqn:BCH}, note that using repeated parameters (e.g. $\alpha = \beta$) will make the coefficients preceding the commutators have the same absolute value (e.g. $\alpha^2 \beta = \alpha \beta^2$). This is evidently unfavorable to generate more linearly independent $\partial_j C (\vec{\vartheta})$ due to the symmetries created. 

In principle, it would be necessary to prove that the expressivity of $C_{seq}^{n+1}$ does not increase for any $\vartheta_{n+1} \in \mathbb{R}$. However, as long as there exists one $\vartheta_{n+1}$ that increases the dimensional expressivity for an operator $\hat{A}_n$, the set of $\{\tilde{\vartheta}_{n+1}\} \subset \mathbb{R}$ that would not raise the expressivity will have measure zero. This can be justified as follows. Assume that the first $n$ parameters are independent (i.e. $\det \left(S_{n}\right) \neq 0$), with $n$ less than the maximal dimensional expressivity, and that there exist some parameters that can increase the expressivity. This implies that the analytic function $f(\Vec{\vartheta}):= \det \left(S_{n+1}\right) $ is not constant 0. The set of parameters that would not increase the expressivity belong to $f^{-1}(0)$. With the regular level set theorem~\cite{lee2012smooth}, $f^{-1}(0)$ is an $n$-dimensional manifold in the $(n+1)$-dimensional parameter space $\mathcal{P}$. Thus, the set of parameters that would not increase the expressivity has Lebesgue measure zero in $\mathcal{P}$. In other words, by choosing $\vartheta_{n+1}$ randomly we increase the dimensional expressivity with probability 1. 
%[a more detailed mathematical argument / reference is needed here]. 

The next section uses these ideas to systematically design quantum circuits that can be used to determine for a controlled quantum system whether it is pure-state controllable or not.

\subsection{Controllability test}

Given a system with operators $\hat{H}_j$ with $0\leq j \leq m$ (cf. Eq.~\eqref{eqn:controlH}), we define the parametric quantum circuit
\begin{equation}
\begin{aligned}\label{eqn:PSC_test}
    C_{PSC}(\vec{\vartheta}) = & \Biggl(\prod_{j=0}^{n_l-1} \hat{R}_{m}(\vartheta_{j(m+1)+m }) ...  \\
        &  \hat{R}_{1}(\vartheta_{j(m+1) +1}) \hat{R}_{0 } (\vartheta_{j(m+1)})\Biggr) \ket{\psi_0},    
\end{aligned}  
\end{equation}
where $\ket{\psi_0}$ is the initial state of the circuit, $m$ the total number of controls in the system and $n_l$ the number of layers in the circuit. A diagram of this circuit is shown in Figure~\ref{fig:PSC_circuit} for a three-qubit example. The initial state $\ket{\psi_0}$, chosen and fixed at the start of the circuit, can be any pure state.  The number of layers $n_l$ should be decided at the start of the algorithm.  All gates in $C_{PSC}(\vec{\vartheta})$ are parametric with different parameters $\vartheta_k$, ranging from $\vartheta_0$ to $\vartheta_{n_l m -1}$. Each of the $n_l$ layers in the circuit has a similar architecture: It starts with the rotation $\hat{R}_0$ around the drift Hamiltonian, an entangling gate if it includes time-independent qubit couplings, and then a sequence of local gates, from $\hat{R}_1$ to $\hat{R}_m$, that use all the different controls sorted by a chosen order.

\begin{center}
\begin{figure}[tb]
\includegraphics[width=0.45\textwidth]{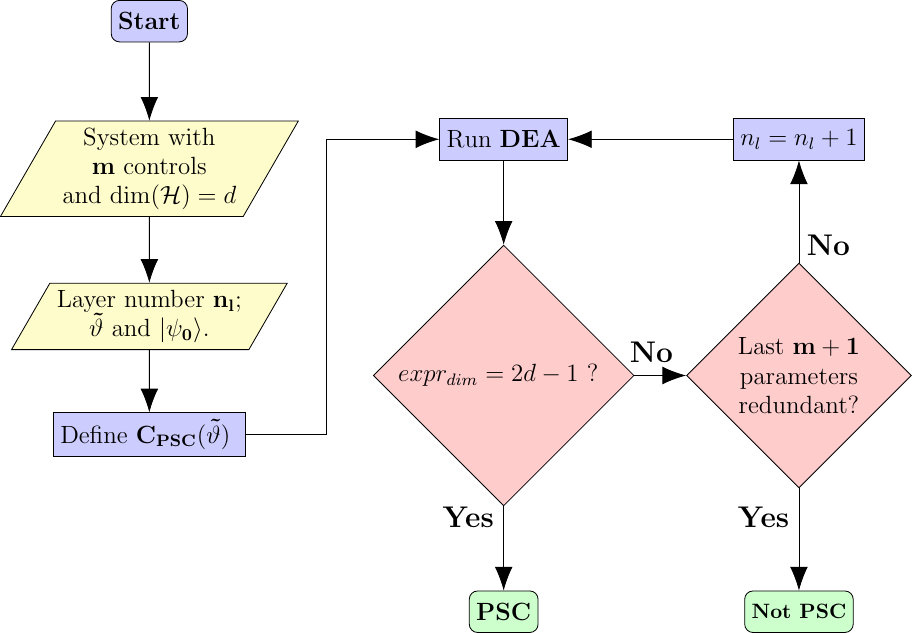}
\caption{\label{fig:flowchart_PSC}  Flowchart for the pure-state controllability algorithm. The yellow rhomboids show the initial inputs necessary to define the circuit $C_{PSC}(\Vec{\vartheta})$. }
\end{figure}
\end{center}

The pure-state controllability test for a system evolving under the Hamiltonian (\ref{eqn:controlH}) is then defined as follows: If the circuit (\ref{eqn:PSC_test}) reaches maximal expressivity, the system is controllable. A schematic flowchart of the pure-state controllability test is shown in Figure~\ref{fig:flowchart_PSC}. If the maximum expressivity of $2d-1$ for a Hilbert space with $\text{dim} (\mathcal{H}) = d$ has not been met with $n_l$ layers, another layer can be added (encompassing a full set of rotation gates with their respective new parameters) and the test can be repeated for the new circuit with $n_l+1$ layers. By definition, the dimensional expressivity can only augment at the rate of one per parameter $\vartheta_j$ at maximum. For a system with $m$ controls, there are a total of $m+1$ parameters per layer. Therefore, the minimum number of layers needed to reach maximum expressivity for $m$ controls is 
\begin{equation}\label{eqn:min_nl}
    n_{l, \, \text{min}} = \left\lceil\frac{2d-1}{m+1}\right\rceil . 
\end{equation}
Since layers may have some redundant parameters, the dimensional expressivity may not necessarily rise at the maximum rate and more layers may have to be included. Consequently, the algorithm is best started with the minimum number of layers required to achieve maximum expressivity and additional layers shall be concatenated as needed. 

It may as well happen that the dimensional expressivity remains the same even with the inclusion of a new layer. In this case the test stops, as the dimensional expressivity will not further increase. In instances where the dimensional expressivity reaches a plateau, it is necessary to double-check using a different array of random parameters $\vec{\vartheta}$ and repeat this comparison with the $n_l$- and $n_l+1$-layered circuits, following the reasoning explained in section~\ref{ssec:expr_and_control}. Using a random set of parameters will yield an answer on whether the expressivity can be increased or not with probability 1. If the dimensional expressivity remains at a value less than $2d-1$ for a sufficiently large set of different random parameters, then the system is labelled not pure-state controllable and the test concludes. 

The algorithm will always end with an affirmative or negative result regarding pure-state controllability. The loop in Figure~\ref{fig:flowchart_PSC} will be exited under one of the following conditions: Either maximal dimensional expressivity is reached or a last layer exclusively composed of redundant parameters is found. In other words, the method ends when the finite upper bound of the dimensional expressivity has been reached or when the expressivity before and after the addition of a new layer remains the same. Since the dimensional expressivity is always an integer, the loop must conclude in a finite number of iterations. 

Parameters with repeated values in the same rotation gates (e.g. $\vartheta_{p} = \vartheta_{q}$ on gates $\hat{R}_{j}(\vartheta_{p})$ and $\hat{R}_{j}(\vartheta_{q})$ for a certain $j$) are usually detrimental to reach maximum expressivity. A trivial example is the case of $\Vec{\vartheta}=\vec{0}$, where the maximum possible dimensional expressivity of $C_{PSC}(\vec{0})$ is always $m+1$, with $m$ the number of local controls. 

\correction{A more detailed description of the algorithm can be found in Appendix \ref{sec:pseudocode}. This includes step-by-step pseudo code and the indication which parts of the method can be performed classically and which parts with quantum computations. }

%    ----    ----    ----    ----    ----    ----    ----- 
\subsection{Examples}
To illustrate the described algorithm, we consider a four-qubit array with the following Hamiltonian: 
\begin{equation}\label{eqn:4qubit} 
    \hat{H}_{4q}(t) = \sum_{j=0}^{3} -\frac{\omega_j}{2}\hat{\sigma}_z^{j} + \sum_{k=0}^{2} J_{k, k+1}\hat{\sigma}_x^{k}\hat{\sigma}_x^{k+1}
      +  \hat{H}_{ctrl} (t) 
\end{equation}
The first term encompasses the free-qubit Hamiltonians and the second one contains the time-independent couplings. The qubit frequencies $\omega_j$ and the coupling strengths $J_{k, k+1}$ have been chosen to fit the ones normally used in superconducting circuits~\cite{kjaergaard2020superconducting} and their exact value can be found in Table~\ref{tbl:4qubit_coef}. The last operator, $\hat{H}_{ctrl} (t)$, contains all the relevant information about the controls, including their number and type. We choose two configurations of controls to study two separate systems with Hamiltonian (\ref{eqn:4qubit}), one that is pure-state controllable and one that is not. 

\begin{table}[tb]
\centering
\begin{tabular}{cccccc}
\toprule
  \multicolumn{6}{c}{Coupling strengths (MHz)} \\
\cmidrule(r){3-5} 
 & & $J_{0,1}$ & $J_{1,2}$ & $J_{2,3}$ &\; \\
 & & $170$ & $220$ & $150$              &\;\\
\midrule
\multicolumn{6}{c}{Qubit frequencies (GHz)} \\
\cmidrule(r){2-5} 
\quad & $\omega_0$ & $\omega_1$ & $\omega_2$ & $\omega_3$ &  \\
\textcolor{white}{.}\;\textcolor{white}{.} & $5.40$ & $5.30$ & $5.42$  & $5.37$                & \\
\bottomrule
\end{tabular}
\caption{Parameters for the Hamiltonian (\ref{eqn:4qubit}). The frequencies and the coupling strengths have been chosen in a range that is common for superconducting circuits. 
}
\label{tbl:4qubit_coef}
\end{table}

\begin{center}
\begin{figure}[tb]
\begin{center}
\includegraphics[width=0.35\textwidth]{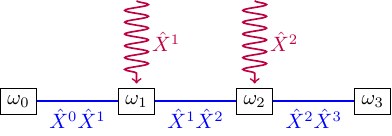}
\end{center}
\caption{\label{fig:PSCtest_PSC} Four-qubit system that is pure-state controllable, cf. Eqs. (\ref{eqn:4qubit}) and (\ref{eqn:4qubit_PSC}).}
\end{figure}
\end{center}

First, we assume the controls from Eq.~\eqref{eqn:4qubit} to be
\begin{equation}\label{eqn:4qubit_PSC} 
    \hat{H}_{ctrl} (t) = u_1 (t) \hat{\sigma}_x^1+  u_2 (t) \hat{\sigma}_x^2.    
\end{equation}
This system is operator controllable, as proven by the Lie algebra rank condition~\cite{dAlessandro2021} and the graph method~\cite{gago2023graph}. This in particular implies that it is also pure-state controllable. A diagram of the system may be found in Figure~\ref{fig:PSCtest_PSC}. 

Since the system only has two controls, each layer of the circuit will have exactly 3 gates—the entangling gate involving the drift and the two related to the local controls coupling to $\hat{\sigma}_x^1$ and $\hat{\sigma}_x^2$, respectively. We have chosen $\ket{\psi_0}=\ket{0000}$ (in the logical basis of the free qubits) as the initial state of the circuit and $n_l = 11$, matching the minimal number of layers to obtain maximum dimensional expressivity (cf. Eq.~\eqref{eqn:min_nl}). For a circuit acting on a four-qubit array, it has a value of $expr_{dim} = 31$. We have generated a random set of parameters $\Vec{\vartheta} \in [0, 2 \pi]^{33}$ (since in this case $(m+1)\cdot n_l = 33$). We have classically simulated the parametric quantum circuit and calculated the $S_{C_{PSC}, n} (\vec{\vartheta})$ matrices from Eq.~\eqref{eqn:Smatrix}. We have both determined the redundant parameters in the circuit and estimated the dimensional expressivity. 

In these simulations, the maximum dimensional expressivity is steadily reached, with every layer raising it by 3. The maximum value of $expr_{dim} = 31$ is achieved with the first parameter of the last layer, proving that the system is pure-state controllable. In this example the minimum number of layers that we had chosen was enough to reach maximum expressivity. The same behaviour has been observed for all the different random sets of parameters $\Vec{\vartheta}$ tested. The same configuration of gates was further tested using different random initial states $\ket{\psi_0}$, yielding similar results. 

%    --    --    --    --    --    --    --    --    --    

%\subsubsection{Four-qubit system that is not pure-state controllable}

\begin{figure}[tb]
\begin{center}
\includegraphics[width=0.35\textwidth]{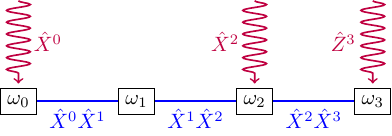}
\end{center}
\caption{\label{fig:PSCtest_notPSC}  Four-qubit system that is not pure-state controllable, cf. equations (\ref{eqn:4qubit}) and (\ref{eqn:4qubit_notPSC}).}
\end{figure}

Second, we present a system that is not pure-state controllable, whose control operators are
\begin{equation} \label{eqn:4qubit_notPSC}   
\hat{H}_{ctrl} (t) = u_1 (t) \hat{\sigma}_x^0+  u_2 (t) \hat{\sigma}_y^2+  u_3 (t) \hat{\sigma}_z^3, 
\end{equation}
cf. Figure~\ref{fig:PSCtest_notPSC}. 
\correction{The dimension of its Lie algebra $\mathcal{L}$ can be found following the method described in Ref.~\cite{dAlessandro2021}. To this end, a basis of $\mathcal{L}$ must be generated, whose cardinality will be equal to the dimension of the Lie algebra. Given a system following Equation (\ref{eqn:controlH}), we can compute a basis by starting with a linearly independent set of the elements of zeroth order: The drift $\hat{H}_0$ and the control operators $\hat{H}_j$ (for $1\leq j \leq m$). We complete the basis by including the nested commutators of the elements of zeroth order that are linearly independent, i.e. $[\hat{H}_{j_1}, [\hat{H}_{j_2}, ...[\hat{H}_{j_k}, \hat{H}_{j_{k+1}}]...]]$. Since the dimension has an upper bound, this method must converge in a finite number of iterations.}
\correction{In the case of Equations (\ref{eqn:4qubit}) and (\ref{eqn:4qubit_notPSC}), we reach a dimension of} $\dim (\mathcal{L})=120 < \dim (\mathfrak{su}(16)) = 255$, which only proves that the system is not operator controllable. The system would be pure-state controllable if and only if 
\begin{equation}
    \dim \left(Lie \left([\rho_0, \mathcal{L}]\right)\right) = 2 \dim(\mathcal{H}) - 2
\end{equation}
with $\rho_0 = \ket{0000}\bra{0000}$~\cite{dAlessandro2021}. We confirm that the system is not pure-state controllable since $\dim \left(Lie \left([\rho_0, \mathcal{L}]\right)\right) = 28 < 30$ for the current system. Even though there are more local controls than in the first example, the system is not controllable due to their positions. Similarly as before, we create a circuit with four gates (related to the drift and the three local controls) per layer. We choose a minimum number of layers $n_l = 8$ (different to the one before due to the different number of controls), $\ket{\psi_0}=\ket{0000}$ and a set of random parameters $\Vec{\vartheta} \in [0, 2 \pi]^{32}$.

At the end of the last layer the dimensional expressivity yields a total of 29 out of the 31 that would imply pure-state controllability. Following the flowchart depicted in Figure~\ref{fig:flowchart_PSC} we have added a new layer ($n_l = 9$) with a new set of random parameters and repeated the dimensional expressivity analysis. According to our simulation, the new layer contains only redundant parameters (i.e. the expressivity remains at 29), which stops the algorithm and means that the system is not pure-state controllable. To verify the validity of this outcome, we have repeated the test for multiple different random sets of parameters. In every instance the same result is reached, which leads to the conclusion that the system is indeed not pure-state controllable, as discussed in section~\ref{ssec:expr_and_control}.

%------------------------------------------------

\section{Operator controllability test using dimensional expressivity analysis}\label{sec:OC_controllability}

Operator controllability is the relevant type of controllability for a qubit array in order to perform all quantum logic gates. Its connection to the dimensional expressivity of a circuit is less evident, since dimensional expressivity is related to the different states that can be reached. The Choi-Jamio\l kowski isomorphism~\cite{choi1975, jamiolkowski1972} allows to bridge the gap with a map between operators on a Hilbert space $\mathcal{H}$ and states in $\mathcal{H}\otimes\mathcal{H}$. It is used, for example, in quantum process tomography, allowing to employ techniques from state tomography to operators~\cite{lloyd2014quantum}.
Similarly, by doubling the number of qubits, we can exploit the channel-state duality between operators in the original system and states in the bipartite extended system for controllability analysis.

\subsection{Lifting pure-state to operator controllability via the Choi-Jamio\l kowski isomorphism}

Let us assume a qubit array with Hamiltonian (\ref{eqn:controlH}) for which we seek to determine operator controllability. This system with Hilbert space $\mathcal{H}$ and dimension $dim(\mathcal{H})=d$ will henceforth be referred to as the original system. We then define a bipartite extended system in $\mathcal{H}\otimes\mathcal{H}$ composed of the original system and the same number of auxiliary qubits. To simplify the argument, we first assume no dynamics over the \correction{auxiliary} qubits. Later we extend our discussion to include some local Hamiltonians on the auxiliary qubits. Given any operator $\hat{O}\in L(\mathcal{H}\otimes\mathcal{H})$, we write $\hat{O}^A$ to indicate that the operator only acts non-trivially on the partition of the original system ($A$), i.e.
\begin{equation}
 \hat{O}^A = \hat{Q} \otimes \mathds{1}_d
\end{equation}
for some operator $\hat{Q}$. Analogously, we write $\hat{O}^{AB}$ for operators that act non-trivially on both partitions (the original system and the auxiliary qubits). Neglecting the local contributions of the \correction{auxiliary} qubits, the Hamiltonian of the extended system is given by
\begin{equation}\label{eqn:CJ_H_NoAncillaH}
    \hat{H}^A(t) = \hat{H}(t; u_1, ... u_m) \otimes \mathds{1}_2^{\otimes q} 
\end{equation}
where $q$ is the number of qubits in the original system.

We assume that the extended system can be prepared in a maximally entangled state, 
\begin{equation}\label{eq:psiME}
    \ket{\psi_{ME}} = \sum_{i = 0}^{d-1} \frac{1}{\sqrt{d}} \ket{e_i} \otimes\ket{e_i},
\end{equation}
where $\{\ket{e_i}\}_0^{d-1}$ is an orthonormal basis of $\mathcal{H}$.

We define the circuit on the extended system
\begin{equation}
\begin{aligned}\label{eqn:OC_circuit_NoAncillaH}
    C_{OC}^A(\vec{\vartheta}) := & \prod_{j=0}^{k} \left(\hat{R}^{A}_{m}(\vartheta_{j(m+1) +m }) ... \right. \\
        & \left. \hat{R}^{A}_{1}(\vartheta_{j(m+1) +1}) \hat{R}^{A}_{0 } (\vartheta_{j(m+1) })\right) \ket{\psi_{ME}}.    
\end{aligned}  
\end{equation}

The rotations $\hat{R}_{k}^{A}(\alpha)$ are given by the drift ($k = 0$) and the control operators ($1 \leq k \leq m$) of the original subsystem: 
\begin{equation} \label{eqn:Rgate_OC_i}
   \hat{R}^{A}_{k} (\alpha):= \exp\left(-i\, \frac{\alpha}{2} \hat{H}_k\otimes \mathds{1}_2^{\otimes q}\right), \quad 0 \leq k \leq m, 
\end{equation}
with $\hat{H}_k$ given in Eq.~\eqref{eqn:controlH}.

\begin{figure}[tb]
\includegraphics[width=0.45\textwidth]{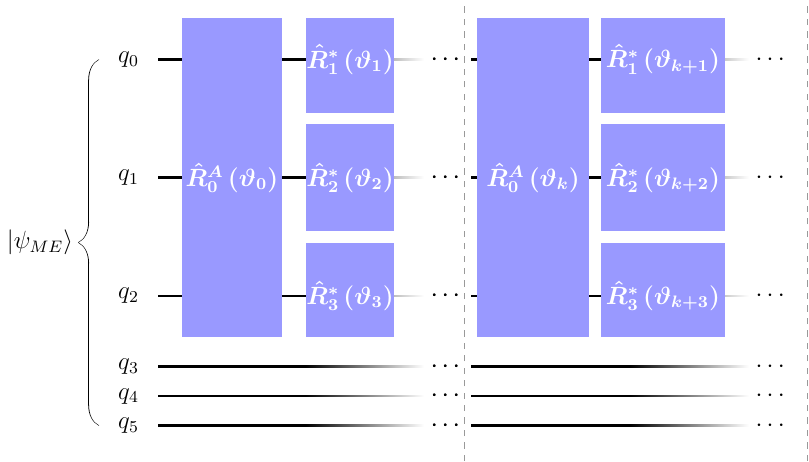}
\caption{\label{fig:OC_circuit_NoAncillaH}  Parametric circuit for the extended system required to perform the operator controllability test (\ref{eqn:OC_circuit_NoAncillaH}) for a three-qubit system. The qubits $q_i$ with $i=0,1,2$ constitute the original system, whereas $q_j$ with $j=3,4,5$ are the \correction{auxiliary} qubits.}
\end{figure}

A visual representation of the circuit is found in Figure~\ref{fig:OC_circuit_NoAncillaH}. The parameter space $\mathcal{P} \ni \vec{\vartheta}$ is assumed to be connected and compact without boundary (e.g. with every coordinate $\vartheta_i$ being cyclic). The final state of the circuit will always be of the form
\begin{equation}\label{eqn:final_state_NoAncillaH}
    C_{OC}^A(\vec{\vartheta}) = \frac{1}{\sqrt{d}} \sum_{i = 0}^{d-1} \ket{e_i} \otimes \left(\hat{U}(\vec{\vartheta})\ket{e_i}\right),
\end{equation}
with $\hat{U}(\vec{\vartheta})$ a unitary operator depending on the circuit's parameters.

Our goal is to prove that dimensional expressivity of the extended system is enough to determine operator controllability of the original system. To this end, we make use of the Choi-Jamio\l kowski isomorphism~\cite{choi1975, jamiolkowski1972, jiang2013channel}. The map it describes is written as
\begin{equation}
\begin{aligned}\label{eqn:CJ_iso}
    \Lambda(\hat{A}) \,:=&\, \left(\mathds{1}_{\mathcal{L}_\mathcal{H}} \otimes \hat{A}   \right) \left(\ket{\phi}\bra{\phi}\right) \\
    \,=&\, \sum_{i,j} \ket{\psi_i} \bra{\psi_j} \otimes \hat{A} \Bigl( \ket{\psi_i}\bra{\psi_j}\Bigr)   
\end{aligned}
\end{equation}
for any operator $\hat{A}$ in the Hilbert space of linear operators on the Liouville space and the unnormalized state $\ket{\phi} = \sum_i \ket{\psi_i} \otimes \ket{\psi_i}$, with $\{\ket{\psi_i}\}_{i = 0}^{d-1}$ an orthonormal basis of $\mathcal{H}$.

Identifying $\hat{A}$ in Eq.~\eqref{eqn:CJ_iso} with $\hat{U}(\vec{\vartheta})$ in Eq.~\eqref{eqn:final_state_NoAncillaH}, we know that 
\begin{equation}
\begin{aligned}\label{eqn:OCtest_iso_density}
    \hat{U}(\mathcal{P}) \, \cong& \,\Lambda(\hat{A}) \\
    =& \,\sum_{i,j = 0}^{d-1} \ket{e_i}\bra{e_j} \otimes \left(\hat{U}(\mathcal{P})\ket{e_i}\bra{e_i}\hat{U}(\mathcal{P})^{\dagger}\right). 
\end{aligned}  
\end{equation}
The operators $\hat{U}(\mathcal{\vec{\vartheta}})$ are unitary for every $\vec{\vartheta}\in \mathcal{P}$, hence purity-preserving. We transform the density matrix representation from Eq.~\eqref{eqn:OCtest_iso_density} into a pure-state representation, resulting in
\begin{equation}\label{eqn:OCtest_iso_A}
    \hat{U}(\mathcal{P}) \cong \sum_{i=0}^{d-1} \ket{e_i} \otimes \hat{U}(\mathcal{P})\ket{e_i}    \cong C_{OC}^A(\mathcal{P}).
\end{equation}
Therefore, there exists an embedding between the evolutions $\hat{U}(\mathcal{P})$ that are generated using a combination of rotations given by the controls and the final states of the circuit $C_{OC}^A(\mathcal{P})$. A system with traceless operators as in Eq.~\eqref{eqn:controlH} and $\dim (\mathcal{H}) = d$ is operator-controllable if and only if the manifold of the unitary evolutions that can be generated $\hat{U}_{\hat{H}}$ is isomorphic to $SU(d)$. Evidently, $\hat{U}(\mathcal{P})\subseteq \hat{U}_{\hat{H}} \subseteq SU(d)$. Since the parameter space $\mathcal{P}$ is connected and compact without boundary, $\hat{U}(\mathcal{P}) = SU(d)$ if and only if $\dim (\hat{U}(\mathcal{P})) = \dim (SU(d))$. Thus, using Eq.~\eqref{eqn:OCtest_iso_A}, the system will be operator-controllable if $\dim (C_{OC}^A(\mathcal{P})) = \dim (SU(d))$, i.e., if the dimensional expressivity of the circuit $C_{OC}^A(\vec{\vartheta})$ is $d^2-1$. 

From here we proceed analogously as the pure-state controllability test from section~\ref{ssec:expr_and_control}. We present the outline of the operator controllability test in Figure~\ref{fig:flowchart_OC}. If the dimensional expressivity is less than $d^2-1$, we inspect the parameters in the last circuit layer. If they all are redundant, the test ends and the system is deemed not controllable. Indeed, if all parameters in the last layer are redundant, we are unable to find more linearly independent operators in the dynamical Lie algebra of the system. If the number of linearly independent elements of the algebra (i.e. number of independent parameters) is less than $\dim (SU(d))$, there exist some unitary operations that cannot be implemented. Therefore, the system is not operator controllable. This step must be checked with multiple arrays of random parameters $\vec{\vartheta}$, as there may be a set of arrays of parameters with measure zero over $\mathcal{P}$ that yield a lower value for the dimensional expressivity. The same arguments we used in section~\ref{ssec:expr_and_control} apply here, as $C_{OC}^A(\mathcal{P})$ is a manifold of states in $\mathcal{H}\otimes \mathcal{H}$. 

If at least one parameter in the last circuit layer is independent, the test continues. We iterate by adding a new layer and calculating the circuit's expressivity. The algorithm will eventually come to an end, either with maximal value for the dimensional expressivity or with a layer of redundant parameters at the end of the circuit.  

\begin{figure}[tb]
\includegraphics[width=0.45\textwidth]{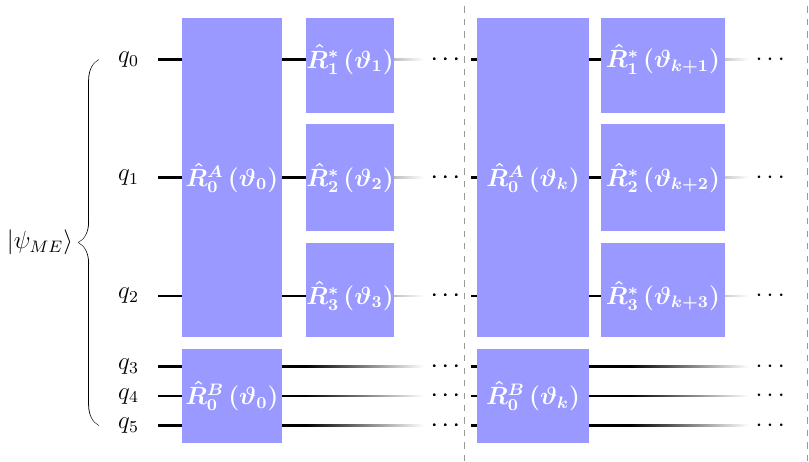}
\caption{\label{fig:OC_circuit_WithAncillaH} Circuit on the extended system required to perform the operator controllability test (\ref{eqn:OC_circuit_WithAncillaH}) for a three-qubit system. The qubits $q_i$ with $i=0,1,2$ constitute the original system, whereas $q_j$ with $j=3,4,5$ are the \correction{auxiliary} qubits. The rotations $\hat{R}_0^{B}$ (cf. Eq.~\eqref{eqn:Rgate_OC_0_B}) include the free-qubit dynamics of the \correction{auxiliary} qubits.}
\end{figure}

We now move to a more realistic setting that incorporates dynamics in the \correction{auxiliary} qubits. We undertake this by including the drift of the auxiliary partition. The new Hamiltonian of the bipartite system is then
\begin{equation}\label{eqn:CJ_H_WithAncillaH}
    \hat{H}^{AB}(t) = \hat{H}(t; u_1, ... u_m) \otimes \mathds{1}_2^{\otimes q} + \sum_{j=0}^{q-1} -\frac{\omega_j}{2}\hat{\sigma}_z^{j + q}, 
\end{equation}
with
\begin{equation}
    \hat{\sigma}_z^{k} := \mathds{1}\otimes ... \otimes \mathds{1}\! \otimes\underbrace{\hat{\sigma}_z}_{\text{$k$ position}}\!\otimes \mathds{1} \otimes ... \mathds{1}.
\end{equation}
It results in the following circuit to test operator controllability
\begin{equation}
\begin{aligned}\label{eqn:OC_circuit_WithAncillaH}
    C_{OC}^{AB}(\vec{\vartheta}) := & \prod_{j=0}^{k} \left(\hat{R}^{A}_{m}(\vartheta_{j(m+1) +m }) ... \hat{R}^{A}_{1}(\vartheta_{j(m+1) +1}) \right. \\
        & \left.  \hat{R}^{B}_{0 } (\vartheta_{j(m+1) })\hat{R}^{A}_{0 } (\vartheta_{j(m+1) })\right) \ket{\psi_{ME}},    
\end{aligned}  
\end{equation}
where
\begin{equation} \label{eqn:Rgate_OC_0_B}
   \hat{R}^{B}_{0} (\alpha):= \exp\left(i\, \frac{\alpha}{2} \, \sum_{j=0}^{q-1} \frac{\omega_j}{2}\hat{\sigma}_z^{j + q}\right). 
\end{equation}
Note that the parameters $\vartheta_{j(m+1)}$ of the gates $\hat{R}^{A}_{0}$ and $\hat{R}^{B}_{0}$ in the same layer $j$ are always the same because there is no active control over these operators---they are due to the time-independent part of the Hamiltonian. In other words, these gates are implemented by letting the system evolve a certain amount of time $t = \vartheta_{j(m+1)}/2$. The number of parameters per layer for a system with $m$ controls remains equal to $m+1$, despite having an extra rotation gate per layer. A diagram of the new circuit is found in Figure~\ref{fig:OC_circuit_WithAncillaH}. 

If we choose an orthonormal basis for the $B$ partition consisting of the eigenstates of the \correction{auxiliary} qubits, then 
\begin{equation}\label{eqn:OCtest_iso_AB}
    C_{OC}^{AB}(\mathcal{P}) \cong \sum_{i=0}^{d-1} \left( \hat{U}(\mathcal{P}) e^{\varphi_i (\Vec{\vartheta})}\ket{e_i} \right)\otimes\ket{e_i}.       .
\end{equation}
The only difference between equations (\ref{eqn:OCtest_iso_A}) and (\ref{eqn:OCtest_iso_AB}) is the local phases $\varphi_i (\Vec{\vartheta})$, which are uniquely determined for any array of parameters $\Vec{\vartheta}$. These do not change the value of the dimensional expressivity since for any array $\Vec{\vartheta}$ there exists a neighborhood in which 
\begin{equation}
     C_{OC}^{A}(\Vec{\vartheta}) \cong  C_{OC}^{AB}(\Vec{\vartheta}).
\end{equation}
This implies the local dimension of the manifold of reachable states to be identical, i.e., the dimensional expressivity to be the same. Therefore, we can include the local Hamiltonians of the \correction{auxiliary} qubits in our calculations to describe a more realistic model and still use the Choi-Jamio\l kovski isomorphism to design the parametric quantum circuit (\ref{eqn:OC_circuit_WithAncillaH}).

\subsection{Controllability test}

\begin{center}
\begin{figure}[tb]
\includegraphics[width=0.45\textwidth]{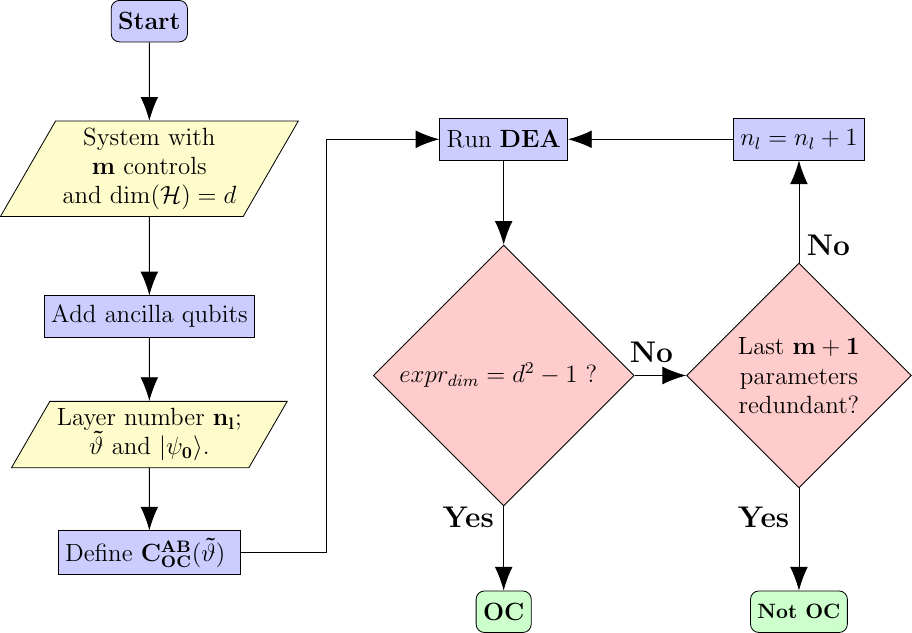}
\caption{\label{fig:flowchart_OC}  Flowchart for the algorithm testing operator controllability. The yellow rhomboids show the initial inputs necessary to define the circuit $C_{OC}^{AB}(\Vec{\vartheta})$.} 
\end{figure}
\end{center}

Once again we consider a qubit array with traceless Hamiltonian (\ref{eqn:controlH}) and the corresponding extended system, composed of the original $q$-qubit array and $q$ more auxiliary qubits. We assume the extra qubits to have arbitrary natural frequencies $\omega_j$, such that the Hamiltonian of the extended system is given by Eq.~\eqref{eqn:CJ_H_WithAncillaH} and the parametric quantum circuit by Eq.~\eqref{eqn:OC_circuit_WithAncillaH}. As shown in Figure~\ref{fig:OC_circuit_WithAncillaH} for a three-qubit example, for a system with $m$ controls the circuit has exactly $m+1$ parameters per layer. As for pure-state controllability, it is encouraged to choose a number of layers $n_l$ that would a priori be sufficient to reach the maximum dimensional expressivity. In the case of operator controllability, it is $\dim (\mathfrak{su}(d)) = d^2 -1$, with $d$ the Hilbert space dimension of the original system, $d = 2^q$ \footnote{We only claim that the value for the maximal dimensional expressivity is $d^2-1$ (with $d=2^q$) for the circuits $C^{AB}_{OC}$ (\ref{eqn:OC_circuit_WithAncillaH}). Other parametric quantum circuits acting on $\mathcal{H} \otimes \mathcal{H}$ could in principle reach higher values of expressivity, up to $2^{2q+1}-1$ }. Thus, the condition for the minimum number of layers to obtain the maximal dimensional expressivity is
\begin{equation}\label{eqn:min_nl_OC}
    n_{l, \, \text{min}} = \left\lceil\frac{d^2-1}{m+1}\right\rceil . 
\end{equation}

With the dimensional expressivity we find the maximum number of linearly independent states in $\mathcal{H} \otimes \mathcal{H}$ that can be generated in a neighborhood of $C_{OC}^{AB}(\vec{\vartheta})$. This in turn yields information about the maximum number of linearly independent operators on $\mathcal{H}$ that can be generated by the original system around the identity. Since we know that these operators belong to the Lie algebra $\mathfrak{su}(d)$ we simply want to determine if we can span all the $d^2 -1$ dimensions in the algebra, i.e. having operator controllability, or not. 

The operator controllability of a system evolving under the Hamiltonian (\ref{eqn:controlH}) is determined as follows: If the circuit (\ref{eqn:OC_circuit_WithAncillaH}) has dimensional expressivity equal to $d^2 -1$, then the system is operator controllable. Analogously to the pure-state controllability test, if this value for the dimensional expressivity is not reached, another layer should be concatenated at the end of the circuit. If all the new parameters in the last layer are redundant, then the system is not operator controllable (with a probability of measure 1); otherwise, the process of concatenating layers shall be repeated. The main steps of the algorithm is displayed in Figure~\ref{fig:flowchart_OC}. Similarly to section~\ref{ssec:expr_and_control}, it is important to to ensure the validity of a result of "not operator controllable" by repeating the test for different arrays of random parameters. 

\correction{Analogous to Section \ref{sec:PSC_controllability}, the pseudo code of the algorithm can be found in Appendix \ref{sec:pseudocode}.}

%    ----    ----    ----    ----    ----    ----    ----- 

\subsection{Examples}

In the following we consider a three-qubit array with Hamiltonian 
\begin{equation}\label{eqn:3qubit} 
    \hat{H}_{3q}(t) = \sum_{j=0}^{2} -\frac{\omega_j}{2}\hat{\sigma}_z^{j} + \sum_{k=0}^{1} J_{k, k+1}\hat{\sigma}_z^{k}\hat{\sigma}_z^{k+1} +  \hat{H}_{ctrl} (t) .
\end{equation}
The second term, containing the time-independent two-qubit couplings, has been modified to $\hat{\sigma}_z^{k}\hat{\sigma}_z^{k+1}$ simply to showcase a qubit interaction different from the one in the previous examples. The qubit frequencies $\omega_j$ and the coupling strengths $J_{k, k+1}$ are listed in  Table~\ref{tbl:3qubit_coef}. We take two different $\hat{H}_{ctrl} (t)$ to study an example that is operator controllable and one that is not. 

\begin{table}[tb]
\centering
\begin{tabular}{cccccc}
\toprule
 & \multicolumn{4}{c}{\makebox[0pt]{Coupling strengths (MHz)}} & \\
\cmidrule(r){2-5}
 & & $J_{0,1}$ & $J_{1,2}$ & &  \\
 & & $170$ & $220$ & & \\
\midrule 
\multicolumn{6}{c}{Qubit frequencies (GHz)} \\
\midrule 
\multicolumn{3}{c}{Original} & \multicolumn{3}{c}{Auxiliary}\\
\cmidrule(r){1-3} 
\cmidrule(r){4-6} 
$\omega_0$ & $\omega_1$ & $\omega_2$ & $\omega_3$& $\omega_4$& $\omega_5$  \\
$5.40$ & $5.30$ & $5.42$  & $5.37$ & $5.29$& $5.34$\\
\bottomrule
\end{tabular}
\caption{Parameters for the Hamiltonian (\ref{eqn:3qubit}) and the \correction{auxiliary} qubits necessary for the circuit (\ref{eqn:OC_circuit_WithAncillaH}).  
}
\label{tbl:3qubit_coef}
\end{table}

\begin{figure}[tb]
\begin{center}
\includegraphics[width=0.3\textwidth]{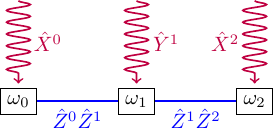}
\end{center}
\caption{\label{fig:OCtest_OC}  Example of a three-qubit system that is operator-controllable, cf. Eq.~\eqref{eqn:3qubit_OC}}
\end{figure}

The first one is given by
\begin{equation}\label{eqn:3qubit_OC} 
    \hat{H}_{ctrl} (t) = u_1 (t) \hat{\sigma}_x^0+  u_2 (t) \hat{\sigma}_y^1 + u_3 (t) \hat{\sigma}_x^2,    
\end{equation}
see Figure~\ref{fig:OCtest_OC}. It is operator controllable as can easily be proven by the Lie algebra rank condition~\cite{dAlessandro2021} and the graph method~\cite{gago2023graph}. 

Since we have 3 controls in the original three-qubit system, the minimum number of layers needed to reach the maximum value of dimensional expressivity for the bipartite system, $expr_{dim} = 63$, is $n_l = 16$ according to Eq.~\eqref{eqn:min_nl_OC}. The orthonormal basis used to define the maximally entangled state $\ket{\psi_{ME}}$ is the logical basis of the free qubits. Last, we generate a random set of parameters $\Vec{\vartheta} \in [0, 2 \pi]^{64}$. Maximum dimensional expressivity of 63 is found for the last parameter of the last layer, confirming that the system is operator controllable.

\begin{figure}[tb]
\begin{center}
\includegraphics[width=0.3\textwidth]{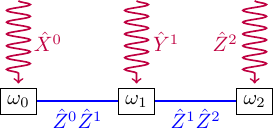}
\end{center}
\caption{\label{fig:OCtest_notOC}  Example of a three-qubit system that is not operator-controllable, cf. Eq.~\eqref{eqn:3qubit_notOC}.}
\end{figure}

For the second example, we choose a different set of controls, 
\begin{equation}\label{eqn:3qubit_notOC} 
    \hat{H}_{ctrl} (t) = u_1 (t) \hat{\sigma}_x^0+  u_2 (t) \hat{\sigma}_y^1 + u_3 (t) \hat{\sigma}_z^2,    
\end{equation}
see Figure~\ref{fig:OCtest_notOC}, making the system not controllable. 
We repeat the same procedure as before, since the number of controls is again $m =3$. At the end of 16 layers the circuit only reaches $expr_{dim } = 31$, which is less than the 63 needed for operator controllability. We could add another layer to verify that every new rotation gate will have a redundant parameter. However, in this case it is sufficient to inspect the rank of the matrices $S_n$ from Eq.~\eqref{eqn:Smatrix} in the last layers. We find that the last independent parameter appears at the end of the tenth layer, with all the remaining ones being exclusively formed by redundant parameters. This is a sufficient condition to determine that the system is not operator controllable (as long as it is verified with multiple sets of random parameters). 

We emphasize that it is important to corroborate every ``not controllable'' result with different arrays $\vec{\vartheta}$ chosen at random. Selecting $\vec{\vartheta}$ in a non-randomized fashion may lead to cases where the dimensional expressivity is lower than the maximum value reached with other different parameters. This would yield wrong results in terms of controllability. It is easily rationalized in terms of symmetries of the commutators $[\hat{H}_i, \hat{R}^A_k(\vartheta_j)]$. These are linked to the partial derivatives of the circuit $\partial_i C_{OC}^{AB}(\vec{\vartheta})$ and to the dimensional expressivity of the circuit. Performing further numerical tests on the previously discussed examples, we have experimented with selecting parameters instead of choosing them at random. Wrong results with lower dimensional expressivity arose when all the parameters were chosen to be the same, e.g. $\vartheta_j = 1$ for every $j$. In every instance, these problems vanished as soon as we generated a new set of random parameters. 

Another important issue concerns the minimum tolerance $\tau$ used to determine the rank of the $S_n$ matrices. More precisely, $\tau$ represents the threshold at which the values of the singular value decomposition of $S_n$ are considered zero. $\tau$ is crucial to determine the different redundant parameters and the expressivity of the circuit. If $\tau$ is too high, then some linearly independent vectors might be deemed dependent by mistake, which would revert on a wrong lower value of the circuit expressivity, potentially turning a controllable system into a fake non-controllable one. Conversely, if $\tau$ is too small some errors might start to add up to make linearly dependent vectors look as if they were independent, falsely showing some parameters as independent. This would in turn raise the dimensional expressivity, usually above the $d^2-1$ threshold that we know to be valid for the case of the operator controllability test. To avoid these cases, it is advisable to use operators with similar orders of magnitude and try different ranges for $\tau$ depending on the order of magnitude of the operators $\hat{H}_j$ from Eq.~\eqref{eqn:controlH}. If the dimensional expressivity analysis is performed on quantum hardware, the tolerance $\tau$ will also depend on the device noise. Indeed, the accuracy of the measurements and the circuit dynamics will take a toll on the accuracy of the rank of the matrices $S_n$. Inevitably, noisier devices will require higher tolerances to determine whether there are redundant parameters (i.e. whether $\det (S_n) =0$) or not.

%------------------------------------------------

\section{Discussion and Conclusions}\label{sec:conclusions}

We have introduced two hybrid quantum-classical algorithms to test pure-state and operator controllability of qubit arrays. As opposed to usual Lie rank and graph methods, the presented algorithms are run directly on a quantum circuit designed to mimic the dynamics of the quantum system to be studied. \correction{The method is also devised as an alternative to the cases where the dynamical Lie algebra can no longer be evaluated analytically or numerically on a classical computer.} We have showcased the capabilities of the procedure with four paradigmatic examples that cover all scenarios for pure-state and operator controllability. 

A useful application of these tests is the resource-efficient design of quantum chips. Our algorithm provides a systematic way to deduce the minimal number of local controls and qubit couplings required to maintain controllability, as a prerequisite of universal quantum computation. In other words, it allows one to identify redundant controls and thus to ease scaling up the quantum chip size. Importantly, the tests allow to obtain this information before the devices are built, as long as the associated quantum circuit can be implemented on a different device. 
Note that while the rank analysis of the $S_n$ matrices scales with the size of the system Hilbert space, this does not pose a fundamental limitation. It can be overcome by mapping the rank computation to a quantum device. More precisely, the quantum device would then be used to find the lowest eigenvalue of $S_n$ in order to determine whether a parameter is redundant or not.
This permits the efficient identification of redundant parameters and the removal of their parametric gates in the circuit. Noise in the device running our hybrid algorithm will limit the accuracy of the lowest eigenvalue and thus determine the minimum threshold for an eigenvalue to be considered zero. 

In addition to its practical aspects, at the conceptual level, our work has revealed the close connection between the controllability of quantum systems and the dimensional expressivity of quantum circuits. In particular, this insight arises from the relation between the states that can be reached in a controllable system and the final states that can be produced in a parametric quantum circuit. The dimensional expressivity analysis allowed us to efficiently quantify the circuit expressivity. Its search for redundant parameters was essential in determining  which controls contributed to reach more states in the Hilbert space. The link between the pure-state and operator controllability test is the inclusion of the Choi-Jamio\l kovski isomorphism that creates a map between operators in a Hilbert space and the states of the extended bipartite space. 

Variational quantum algorithms have previously been used to improve the design of optimal pulses in quantum systems~\cite{magann2021pulses}.  Quantum optimal control theory in general~\cite{GlaserEPJD15,KochEPJQT22} encompassses both the design of the pulse shapes, i.e., control synthesis, and controllability analysis. The controllability tests described here thus extend the use of parametric quantum circuits to the second pillar of quantum optimal control. Quantum optimal control is also closely related to system characterization where controls can be interleaved with free evolutions~\cite{wittler2021integrated, lu2023learning} or applied continuously~\cite{dutkiewicz2023advantage}.

In future work, it will be interesting to study systems with non-local controls, e.g. tunable two-qubit couplings. Moreover, it may be possible to expand our approach to systems other than qubit arrays. To this end, the key task will be to find a mapping from the non-qubit system to the associated quantum circuit that runs on a qubit array. The problem of mapping certain dynamics to a quantum circuit has already been a subject of extensive research, for example, when using parametric variational algorithms for calculating the electronic structure of molecules~\cite{xia2020qubit, kandala2017hardware} or their quantum dynamics~\cite{ollitrault2021molecular}.
\correction{An interesting future perspective is to explore the extension of our approach to the controllability of subgroups. This is sometimes referred to as $G$-controllability, where $G$ is a subgroup of the unitary group $U(n)$. This would be relevant both to open quantum system control and machine learning. While it is not straightforward in the general case, our method can likely be used to analyse certain cases of $G$-controllability but dedicated work in this direction is needed to give a more definitive answer.}
Finally, an intriguing question is how the removal of redundant controls affects the minimum time at which certain dynamics can be implemented, i.e., the quantum speed limit of the system. A controllable system with a new control added can have the same or a lower minimum time for a state transfer or unitary gate. Conversely, removing redundant controls might incur a higher minimum time. Most likely, quantum device design will have to balance the requirements for controllability and operation speed.

%------------------------------------------------

\section*{Acknowledgments}

We gratefully acknowledge financial support from the Einstein Research Foundation (Einstein Research Unit on Near-Term Quantum Devices) and the Deutsche Forschungsgemeinschaft (DFG, German Research Foundation) – Projektnummer 277101999 – TRR 183 (project C05).

\bibliographystyle{quantum}
\bibliography{main}
%\bibliography{main}% Produces the bibliography via BibTeX.

%------------------------------------------------
\clearpage

\onecolumngrid   
\appendix
\section{\correction{Outline of the algorithm}}\label{sec:pseudocode}

%%%%%%%% ALGORITHM Controllability %%%%%%%%%%%%%%%%%%%%%%%%%%%%
\begin{algorithm}[tbp]
\textcolor{gray}{\textbackslash \textbackslash This method is defined for a circuit $C(\Vec{\vartheta})$ only containing parametric rotation gates $\hat{R}_j(\vartheta_j)$}
\newline \textcolor{gray}{\textbackslash \textbackslash $\hat{R}_j(\vartheta_j) :=  \exp\left(-i \,\frac{\vartheta_j}{2} \hat{G_j}\right)$} 
\newline \textcolor{gray}{\textbackslash \textbackslash $ C(\vec{\vartheta}) := \hat{R}_{len(\vec{\vartheta})}(\vartheta_{len(\vec{\vartheta})})\cdots\hat{R}_0(\vartheta_0) \ket{\psi_0}$} 
\newline  
    Input: \newline
        $\bullet$ \textbf{\textit{test\_type}}: it can be 'PSC' or 'OC' depending on which controllability test should be run. 
        \newline  
        $\bullet$ \textbf{\textit{para}}: array $\vec{\vartheta}$ with all parameters $\vartheta_j$. 
        \newline
        $\bullet$ $\mathbf{\hat{G}_{list}}$: list including all the operators $\hat{G}_j$ in matrix form. 
        \newline
        $\bullet$ $\mathbf{\ket{\psi_0}}$: initial state of the circuit.
        \newline
        $\bullet$ \textbf{\textit{last\_lay}}: parameter index at which the last layer starts
        \newline \textcolor{gray}{\textbackslash \textbackslash I.e. the last layer starts with $\hat{R}_{last\_lay}(\vartheta_{last\_lay})$}
        \newline
        $\bullet$ \textbf{\textit{tol}}: tolerance for computing the matrix rank function.

\textit{hildim} $\leftarrow$ len($\ket{\psi_0}$)

\textcolor{gray}{\textbackslash \textbackslash Expressivity right before the last layer}
\newline \textit{expr\_bll} $\leftarrow$ Algorithm\_\ref{alg:classic_DEA}(\textit{para}, $\hat{G}_{list}$, $\ket{\psi_0}$, \textit{last\_lay -1}, \textit{tol})

\textcolor{gray}{\textbackslash \textbackslash Expressivity of the total circuit C}
\newline \textit{expr\_tot} $\leftarrow$ Algorithm\_\ref{alg:classic_DEA}(\textit{para}, $\hat{G}_{list}$, $\ket{\psi_0}$, len(\textit{para}), \textit{tol})

\eIf{\textit{test\_type} is 'PSC'}
    {\textit{max\_exp} $\leftarrow$  $2 hildim -1$
    }{
    \textit{max\_exp} $\leftarrow$  $ hildim^2 -1$    
    } 

\eIf{ $expr\_tot \geq max\_exp$}
    {
        \textit{test\_result} $\leftarrow$ 1 
        \newline \textcolor{gray}{\textbackslash \textbackslash System is controllable}
    }{\eIf{ $expr\_tot > expr\_bll$}
          {
              \textit{test\_result} $\leftarrow$ 2 
              \newline \textcolor{gray}{\textbackslash \textbackslash Test is inconclusive. Repeat test for a circuit containing an additional layer}
          }{
              \textit{test\_result} $\leftarrow$ 0 
              \newline \textcolor{gray}{\textbackslash \textbackslash System is not controllable}
              }  
    } 

    Output: \newline
        $\bullet$ \textbf{\textit{expr\_tot}}: circuit dimensional expressivity
        \newline
        $\bullet$ \textbf{\textit{test\_result}}: 0, 1 or 2 depending on whether the system is not controllable, controllable or the test is inconclusive. For inconclusive tests, one can repeat the algorithm adding a new layer to the circuit.  

  \caption{\correction{Controllability test using dimensional expressivity analysis}}
  \label{alg:controllability}
\end{algorithm}
%%%%% END OF ALGORITHM Controllability %%%%%%%%%%%%%%%%%%%%%%%%

\correction{This appendix includes an outline for the methods described in Figures \ref{fig:flowchart_PSC} and \ref{fig:flowchart_OC} in the form of pseudo code. Algorithm \ref{alg:controllability} displays the main steps for applying the dimensional expressivity analysis to a circuit for the pure-state controllability test as defined in Equation (\ref{eqn:PSC_test}) and the operator controllability test given in Equation (\ref{eqn:OC_circuit_WithAncillaH}). There are two main differences between the two cases: The circuit definition (including the rotation gates $\hat{R}_j(\alpha)$ and the initial state $\ket{\psi_0}$) and the maximum dimensional expressivity $expr_{dim}$ that the circuit has to reach to determine whether the system is controllable or not.}

\correction{For a pure-state controllability test, one must set $test\_type \, = \, 'PSC'$. The circuit description is passed in terms of a list of parameters \textit{para} ($\Vec{\vartheta}$ from Equation (\ref{eqn:PQC})), a list of operators for the rotation gates $\hat{G}_{list}$ (given by the drift and the control operators, cf. Equation (\ref{eqn:Rgate_PSC})), the initial state $\ket{\psi_0}$ (which can be chosen freely) and the parameter index \textit{last\_lay} at which the last circuit layer starts. A numerical tolerance \textit{tol} is also required for computing the rank of the matrices $S_{C, n}$ (cf. Equation (\ref{eqn:Smatrix}).}

\correction{For the operator controllability test, $test\_type$ should be $'OC'$. The circuit should be defined including the auxiliary qubits, as depicted in Figure \ref{fig:OC_circuit_WithAncillaH}. This encompasses the definition of the generators of rotations which are passed as $\hat{G}_{list}$. In this case, the initial state $\ket{\psi_0}$ must be the maximally entangled state $\ket{\psi_{ME}}$ shown in Equation (\ref{eq:psiME}). The rest of the inputs are treated analogously to the previous case.}

\correction{Finally, for either type of controllability, the computation of the $S_{C, n}$ matrices can be done with classical numerical calculations (as shown in Algorithm \ref{alg:classic_DEA}) or it may be achieved using real quantum circuits, as seen in \cite{funcke2021DEA}. This manuscript showcases examples using the former one, although the latter is the intended version for the devised hybrid quantum-classical controllability test.}

%%%%%%%% ALGORITHM Sn without DEA %%%%%%%%%%%%%%%%%%%%%%%%%%%%
\begin{algorithm}[tbp]
\textcolor{gray}{\textbackslash \textbackslash This method is defined for a circuit $C(\Vec{\vartheta})$ only containing parametric rotation gates $\hat{R}_j(\vartheta_j)$}
\newline \textcolor{gray}{\textbackslash \textbackslash $\hat{R}_j(\vartheta_j) :=  \exp\left(-i \,\frac{\vartheta_j}{2} \hat{G_j}\right)$} 
\newline \textcolor{gray}{\textbackslash \textbackslash $ C(\vec{\vartheta}) := \hat{R}_{len(\vec{\vartheta})}(\vartheta_{len(\vec{\vartheta})})\cdots\hat{R}_0(\vartheta_0) \ket{\psi_0}$} 
\newline  
    Input: \newline  
        $\bullet$ \textbf{\textit{para}}: array $\vec{\vartheta}$ with all parameters $\vartheta_j$. 
        \newline
        $\bullet$ $\mathbf{\hat{G}_{list}}$: list including all the operators $\hat{G}_j$ in matrix form. 
        \newline
        $\bullet$ $\mathbf{\ket{\psi_0}}$: initial state of the circuit.
        \newline
        $\bullet$ \textbf{\textit{n}}: Dimension of the square matrix $S_n$ to be calculated  (cf. Eq. (\ref{eqn:Smatrix})).
        \newline
        $\bullet$ \textbf{\textit{tol}}: tolerance used for computing matrix rank function.

\textit{hildim} $\leftarrow$ len($\ket{\psi_0}$)

$J_n$ $\leftarrow$ zero\_array[2\textit{hildim},\textit{n}] %\textcolor{gray}{\textbackslash \textbackslash $\text{hildim} \times n$ null array }

\For{$j$ in $1, ..., n$}{
\textit{$\partial C_j$} $\leftarrow$ $   \frac{\partial C}{\partial \vartheta_j} (\vec{\vartheta})$

$J_n[0:hildim, j-1]$ $\leftarrow$ $Re \left(\partial C_j \right)$

$J_n[hildim: 2 hildim, j-1]$ $\leftarrow$ $Im \left(\partial C_j \right)$
}

\textit{Sn} $\leftarrow$ $J_n^T J_n$

\textit{rank\_Sn} $\leftarrow$ $\rank ( Sn, tol)$ \textcolor{gray}{\textbackslash \textbackslash compute rank of matrix \textit{Sn} with tolerance \textit{tol}}

    Output: \newline
        $\bullet$ \textbf{\textit{rank\_Sn}}: rank of the matrix $S_{C,n}$ 

  \caption{\correction{Classical calculation of $\rank (S_{C,n})$ (cf. Eq. (\ref{eqn:Smatrix})). \newline This classical algorithm can be replaced by measurements on a quantum device as defined in \cite{funcke2021DEA}.}}
  \label{alg:classic_DEA}
\end{algorithm}
%%%%% END OF ALGORITHM Sn without DEA %%%%%%%%%%%%%%%%%%%%%%%%

\end{document}